\documentclass[%
 aip,
 amsmath,amssymb,
 reprint,%
]{revtex4-1}

\usepackage{graphicx}% Include figure files
\usepackage{dcolumn}% Align table columns on decimal point
\usepackage{bm}% bold math
\usepackage[english]{babel}
\usepackage[utf8]{inputenc}
\usepackage[T1]{fontenc}
\usepackage{amsmath}
\usepackage{mathptmx}
\usepackage{color}
\usepackage{siunitx}
\usepackage{color}
\usepackage{multirow}
\usepackage{ifthen}
\usepackage{bbm}

\definecolor{orange}{rgb}{1,0.5,0}
\definecolor{darkgreen}{rgb}{0,0.4,0.1}

\newcommand{\highlightrevision}{false}

\ifthenelse{\equal{\highlightrevision}{true}}
{

\newcommand{\revbar}[1]{{\color{red}{\sout{#1}}}}
}
{

\newcommand{\revbar}[1]{}
}

\newcommand*\bfr{{\bf r}}
\newcommand*\bfrN{{\bf r}^N}
\newcommand*\bfp{{\bf p}}
\newcommand*\bfpN{{\bf p}^N}
\newcommand*\bff{{\bf f}}
\newcommand*\bfF{{\bf F}}

\newcommand*\bfR{{\bf R}}

\begin{document}

\preprint{AIP/123-QED}

\title{
Use the force!
Reduced variance estimators for densities, radial distribution functions 
and local mobilities in molecular simulations
}

\author{Benjamin Rotenberg}
\affiliation{\small Sorbonne Universit\'es, CNRS, Physico-Chimie des \'electrolytes et
Nanosyst\`emes Interfaciaux, F-75005 Paris, France}
\email{benjamin.rotenberg@sorbonne-universite.fr}

\markboth{}% 
{}

\date{\today}

\begin{abstract}
Even though the computation of local properties, such as densities or radial
distribution functions, remains one of the most standard goals of molecular
simulation, it still largely relies on straighforward histogram-based
strategies. Here we highlight recent developments of alternative approaches leading, 
from different perspectives, to estimators with a reduced variance compared to
conventional binning. They all make use of the force acting on the particles, 
in addition to their position, and allow to focus on the non-trivial part
of the problem in order to alleviate (or even remove in some cases) the
catastrophic behaviour of histograms as the bin size decreases. The corresponding 
computational cost is negligible for molecular dynamics simulations,
since the forces are already computed to generate the configurations, and the
benefit of reduced-variance estimators is even larger when the cost of
generating the latter is high, in particular with \emph{ab initio} simulations.
The force sampling approach may result in spurious residual non-zero values
of the density in regions where no particles are present, but strategies are
available to mitigate this artefact. We illustrate this approach on 
number, charge and polarization densities,
radial distribution functions and local transport coefficients, discuss the
connections between the various perspectives and suggest future challenges for
this promising approach.
\end{abstract}

\maketitle

%%%%%%%%%%%%%%%%%%%%%%%%%%%%%%%%%%%%%%%%%%%%%%%%%%%%%%%%%%%%%%%%%%%%%%%%%%
%%%%%%%%%%%%%%%%%%%%%%%%%%%%%%%%%%%%%%%%%%%%%%%%%%%%%%%%%%%%%%%%%%%%%%%%%%
\section{Introduction}

Molecular Dynamics (MD) and Monte Carlo (MC) algorithms allow to sample
configurations from a statistical ensemble and to numerically compute 
observable properties as averages over configurations for realistic models
of interacting atoms or molecules, when analytical theories are usually 
limited to cruder models~\cite{hansen_theory_2013}.
Since the early days of simulations, the tremendous development of accurate 
quantum and classical descriptions and of efficient sampling algorithms,
supported by the ever growing availability of computational resources,
established molecular simulation as an essential tool and,
more fundamentally, a new scientific approach to investigate the properties of 
Matter~\cite{battimelli_computer_2020}.

The determination of the local properties in condensed matter 
has always been one the main applications of molecular simulation.
To mention but a few examples in Biology, Chemistry and Physics,
one can cite the characterization of the local
structure: of water around solutes such as biological molecules by 3D
densities~\cite{Abel2007,altan_learning_2018,wall_biomolecular_2019};
of interfacial number and charge densities at electrochemical
interfaces involving aqueous electrolytes or room temperature ionic 
liquids (RTIL)~\cite{josesegura_adsorbed_2013,merlet_electric_2014,
kornyshev_three-dimensional_2014,elbourne_nanostructure_2015};
of the complex solvation structure in RTILs~\cite{dommert_comparative_2008}
or of aqueous solutes near an electrode surface~\cite{limmer_nanoscale_2015};
of the radial distribution in concentrated electrolytes~\cite{coles_correlation_2020} 
in order to understand the intriguing behavior of the correlation length
reported in these systems~\cite{smith_electrostatic_2016}.
These structural properties are also used to infer thermodynamic
quantities, such as local solvation entropies, energies or free
energies around proteins~\cite{Nguyen2012}, or Kirkwood-Buff integrals in
mixtures~\cite{kirkwood_statistical_1951}. Finally, the densities or radial distribution 
functions obtained from molecular simulations are often used as reference data to test 
and/or parameterize liquid state theories such as 3D-RISM~\cite{Yoshida2009,Stumpe2011} 
or molecular density functional 
theory~\cite{Ding2017,jeanmairet_molecular_2019,Zhao2011,Jeanmairet2013} as well as 
coarse-grained models~\cite{lyubartsev_calculation_1995,chaimovich_coarse-graining_2011} 
for mesoscale simulations.

Compared to the development of models to describe the systems and of algorithms to sample 
statistical ensembles, surprizingly little effort has been devoted to the improvement of
estimators to compute local properties such as densities or radial distribution 
functions from the available configurations. In practice, one generally relies
on histograms couting the number of particles in a voxel of finite size $h^d$
(in $d=1$, 2 or 3 dimensions) around a given point or of pairs separated by a 
distance comprised between $r$ and $r+\Delta r$. As discussed in more detail
below, these straightforward estimators provide the correct expectation value,
but behave poorly when increasing the resolution, with a diverging variance as 
$h$ or $\Delta r\to0$. The benefit of estimators with a reduced variance would 
be even larger when the computational cost to generate configurations is high, 
for example for \emph{ab initio} MD.

Here we highlight recent developments of alternative approaches leading, 
from different perspectives (\emph{e.g.} inspired from zero-variance estimators 
developed in Quantum Monte Carlo~\cite{assaraf_zero-variance_1999,
toulouse_zero-variance_2007,assaraf_improved_2007}, or directly rooted 
in Statistical Mechanics~\cite{schultz_reformulation_2016}), 
to estimators with a reduced variance compared to conventional binning for densities 
or radial distribution functions~\cite{adib_unbiased_2005,borgis_computation_2013,
de_las_heras_better_2018,purohit_force-sampling_2019,coles_computing_2019},
or even local transport coefficients~\cite{mangaud_sampling_2020}. 
A common point of all the resulting expressions is to use the force acting 
on the particles, in addition to their position.
In Section~\ref{sec:force}, we introduce the main problem of histogram-based
estimators and the idea behind force sampling strategies.
Section~\ref{sec:density} then illustrates the variance reduction
obtained from force-based estimators on local number densities in several dimensions,
and extensions to charge and polarization densities, or to take into account
constraints for rigid molecules. Sections~\ref{sec:rdf} and~\ref{sec:transport} 
are devoted to radial distribution functions and local transport coefficients,
respectively. The link between these and related approaches is further discussed in
Section~\ref{sec:discussion}, while Section~\ref{sec:conclusion} offers some
suggestions of future challenges.

%%%%%%%%%%%%%%%%%%%%%%%%%%%%%%%%%%%%%%%%%%%%%%%%%%%%%%%%%%%%%%%%%%%%%%%%%%
%%%%%%%%%%%%%%%%%%%%%%%%%%%%%%%%%%%%%%%%%%%%%%%%%%%%%%%%%%%%%%%%%%%%%%%%%%
\section{Force sampling}
\label{sec:force}

\subsection{What's wrong with binning?}
\label{sec:force:binning}

The computation of local densities or radial distribution functions (rdf)
from molecular simulations is based on their statistical mechanical
definitions as ensemble averages over microscopic configurations.
Let us first consider for simplicity a system of $N$ independent atoms,
 without distance or angular constraints to define rigid
molecules (see section~\ref{sec:density:extensions}). A point
in phase space is defined by the set of all positions $\bfrN=\{\bfr_1,\dots,\bfr_N\}$
and momenta $\bfpN=\{\bfp_1,\dots,\bfp_N\}$. The potential energy 
of the system, $U(\bfrN)$, includes the interactions between atoms
and the effect of external potentials. In the canonical ensemble,
with fixed volume $V$ and temperature $T$, the local number density
in particles of type $a$ is defined as~\cite{hansen_theory_2013}
\begin{align}
\label{eq:densitydef}
\rho_a(\bfr) &= \left\langle 
\sum_{i=1}^{N_a} \delta( \bfr_i - \bfr) \right\rangle
\; ,
\end{align} 
where the sum runs over atoms of type $a$, $\delta$ is the (3D) Dirac delta function
and $\left\langle \dots \right\rangle$ denotes an average in the canonical
ensemble. Specifically, the average of an observable $f$ is 
\begin{align}
\label{eq:averagedef}
\left\langle f \right\rangle &=
\frac{1}{\mathcal{Z}} \int {\rm d}\bfrN {\rm d}\bfpN \, f(\bfrN,\bfpN) 
e^{-\beta \mathcal{H}(\bfrN,\bfpN)}
\; ,
\end{align}
where $\beta^{-1}=k_BT$ is the thermal energy, 
$\mathcal{H}(\bfrN,\bfpN)=U(\bfrN)+K(\bfpN)$,
with $K(\bfpN)$ the kinetic energy, is the Hamiltonian of the system,
and $\mathcal{Z}$ is the partition function.
Expressions similar to Eq.~\ref{eq:densitydef} can be written for the 1D- and 2D-densities
in cartesian coordinates, using \emph{e.g.} $\delta(z_i-z)$
and $\delta(x_i-x)\delta(y_i-y)$.
In addition, rdfs can be defined as
\begin{align}
\label{eq:rdfdef}
g_{ab}(r) &= \frac{V}{N_a N_b} \frac{1}{4\pi r^2}
\left\langle \sum_{i=1}^{N_a} 
\sideset{}{'}\sum_{j=1}^{N_b} \delta( r_{ij} - r) \right\rangle
\; ,
\end{align} 
where $r_{ij}$ is the distance between particles $i$ and $j$
and the prime in the second sum indicates that $j=i$ should be excluded
when $b=a$.

These expressions lead straightforwardly to algorithms based on
a discretization of space into a grid of voxels with size $h^d$,
with $h$ the grid spacing and $d\in\{1,2,3\}$, for the density
and similarly of distances, with a bin width $\Delta r$, for the rdfs. 
Histograms are obtained by simply counting the number of particles
in a given voxel (for densities) or of particle pairs separated by
a distance comprised between $r$ and $r+\Delta r$ (for rdfs)
for each configuration, summing over particles or pairs,
and averaging over a collection of configurations generated according to their
weight in the canonical ensemble using MD or MC simulations.

In order to illustrate the limitations of this histogram approach,
which remains the most commonly used to date, let us consider a system
of identical non-interacting particles in 3 dimensions. In the absence of 
an external potential, one expects the density to be uniform and equal to the average
density $\rho=N/V$. If the bin size $h$ is small, each voxel of the grid
discretizing space will be either empty or contain 1 particle, leading
to an instantaneous estimate of the density $\tilde{\rho}$ in this voxel of $0/h^3$
or $1/h^3$, respectively. In addition, the fraction of occupied voxels
for each configuration, which for a sufficiently large number of configurations 
is also the fraction of configurations in which a given voxel is occupied,
is $\alpha=\rho h^3\ll1$. The estimator $\tilde{\rho}$ of the local density
provides the correct average over many configurations, 
since $\left\langle\tilde{\rho}\right\rangle=
\alpha\times h^{-3}+(1-\alpha)\times 0=\rho$. 
The quality of this estimator can be measured by computing its variance:
$\left\langle\delta\tilde{\rho}^2\right\rangle=
\alpha\times(h^{-3}-\rho)^2+(1-\alpha)\times (0-\rho)^2=
\alpha(1-\alpha)h^{-6}\approx\rho/h^3$.
As a result, the variance diverges as the bin size $h$ decreases.

\begin{figure}[ht!]
\includegraphics[width=\columnwidth]{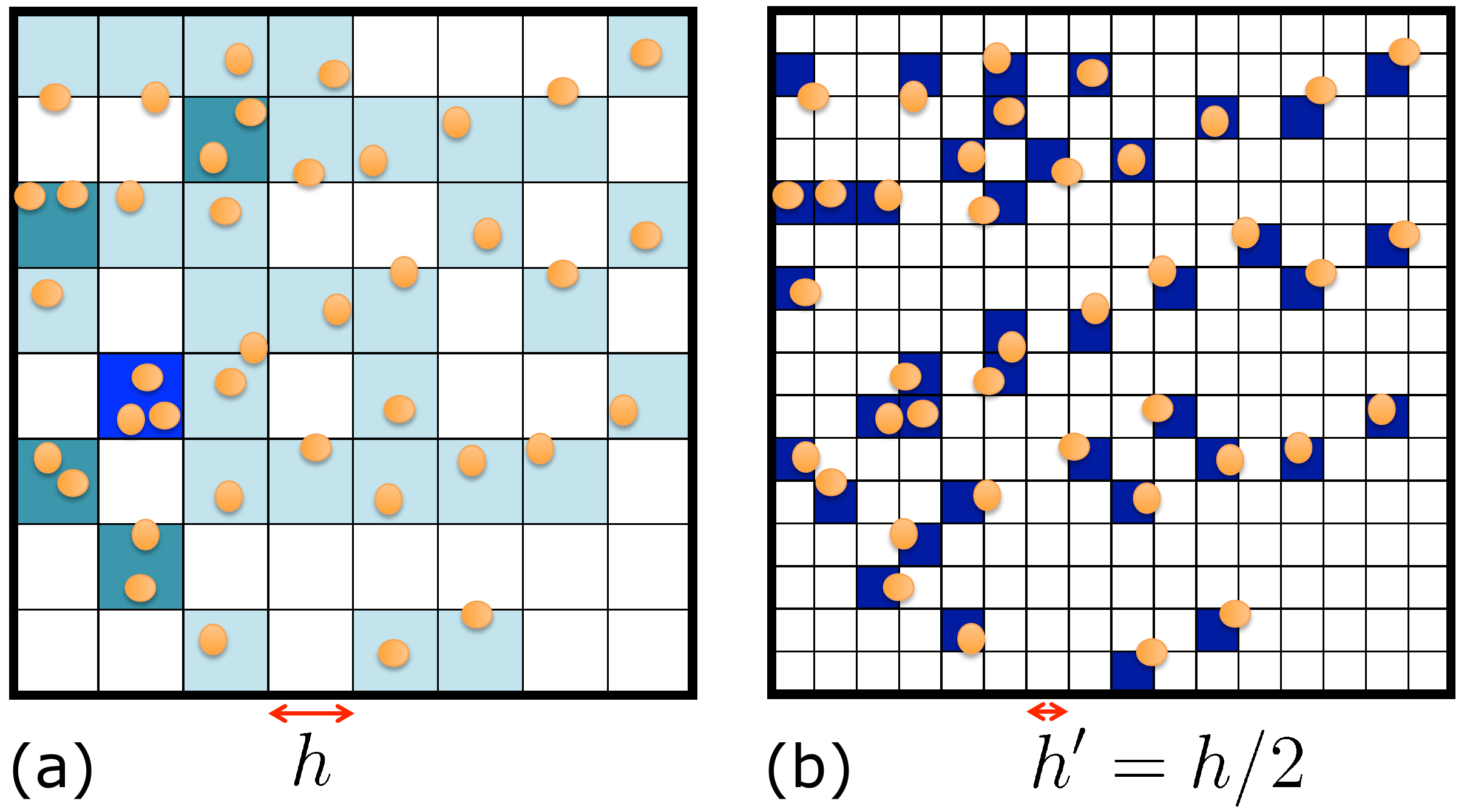}
\caption{
Contribution of a microscopic configuration, with orange disks representing
atoms in a simulation box, to the number density on a grid, estimated using
histograms with bins of width $h$ (a) and $h'=h/2$ (b).
The colors indicate the instantenous estimate of the local density,
from the number of atoms present in each cell and the volume of the latter.
}
\label{fig:binning}
\end{figure}

This problem of histrogram-based estimates of the density 
is illustrated in Figure~\ref{fig:binning}: When the bin size is small,
most voxels are empty and the estimate of the local density is given
by the fraction of configurations in which a given bin is occupied, which is
small and therefore requires a large number of configurations to converge.
Since the computational cost essentially comes from generating the configurations
(and not from the estimate of microscopic properties from these configurations),
improved estimators with a reduced variance would decrease the number of
samples necessary to converge the results. The advantage would be even greater
for computationally intensive approaches, based \emph{e.g.} on \emph{ab initio}
calculations, with which data is usually scarce.

%%%%%%%%%%%%%%%%%%%%%%%%%%%%%%%%%%%%%%%%%%%%%%%%%%%%%%%%%%%%%%%%%%%%%%%%%%
%%%%%%%%%%%%%%%%%%%%%%%%%%%%%%%%%%%%%%%%%%%%%%%%%%%%%%%%%%%%%%%%%%%%%%%%%%
\subsection{The idea behind force sampling}
\label{sec:force:idea}

The previous discussion shows that the variance issue with histograms arises 
essentially from the ideal gas contribution. Fortunately, this contribution is
known: It is simply the average density $\rho=N/V$. In the presence of 
an external potential, the answer is slightly more subtle, since the volume
occupied by the fluid is not known \emph{a priori}. However, the point is that
what really matters are the \emph{variations} in density with respect
to the position (or in rdfs with distance), which result from 
the interactions between the particles and from external potentials.
The idea behind force-sampling strategies is to sample these
variations (the gradients) using estimators involving the force acting 
on the atoms -- a strategy well summarized in the title of
Ref.~\citenum{de_las_heras_better_2018}: ``Better than counting: density profiles
from force sampling''.
In some cases, determining the density (or rdf) from its gradient
numerically requires introducing some discretization. In others, this can be done
analytically, so that the density or rdf can be determined with arbitrary
resolution -- in strong contrast with the conventional histrogram-based methods.

%%%%%%%%%%%%%%%%%%%%%%%%%%%%%%%%%%%%%%%%%%%%%%%%%%%%%%%%%%%%%%%%%%%%%%%%%%
%%%%%%%%%%%%%%%%%%%%%%%%%%%%%%%%%%%%%%%%%%%%%%%%%%%%%%%%%%%%%%%%%%%%%%%%%%
\section{Density}
\label{sec:density}

We begin our survey of improved estimators using force sampling
on the case the number density in Section~\ref{sec:density:number},
highlighting some pitfalls and how to mitigate them in
Section~\ref{sec:density:pitfalls},
before turning to recent extensions to rigid bodies or generic densities
in Section~\ref{sec:density:extensions}.

%%%%%%%%%%%%%%%%%%%%%%%%%%%%%%%%%%%%%
\subsection{Variance reduction from force sampling}
\label{sec:density:number}

Starting from the definition of the density as an ensemble average,
Eq.~\ref{eq:densitydef}, and by differentiating with respect to the
position $\bfr$, one obtains (omitting the species type $a$)
\begin{align}
\label{eq:densitygradient}
\nabla\rho(\bfr) &= \beta \bfF(\bfr)
= \beta
\left\langle  \sum_{i=1}^{N} \delta( \bfr_i - \bfr) \bff_i \right\rangle
\; ,
\end{align} 
where we have introduced the force density $\bfF(\bfr)$ expressed as 
an ensemble average of the force $\bff_i$ acting on the particles, including 
external potentials and the interactions with other particles.
The appearance of the force is due to the gradient of the Boltzmann weight,
since $\nabla_{\bfr_i} e^{-\beta\mathcal{H}} =
e^{-\beta\mathcal{H}} ( -\beta \nabla_{\bfr_i} U ) = 
e^{-\beta\mathcal{H}} \beta \bff_i $ in the average Eq.~\ref{eq:averagedef},
and the properties of the Dirac delta function.
The force sampling approach hence consists in first sampling the force density
from the trajectory, and then computing the density from its gradient.
More precisely, this ``inversion of the gradient'' provides the density
up to a constant, which can be taken as the average density $\rho_0$
to obtain the excess density $\Delta\rho(\bfr)= \rho(\bfr) - \rho_0$.
Such an approach avoids the computation of the ideal gas contribution 
and focuses on the non-trivial contribution from interactions.
In 2 or 3 dimensions, the gradient can formally inverted
as~\cite{de_las_heras_better_2018}
\begin{align}
\nabla^{-1} \cdot \bfF(\bfr) &= \frac{1}{c_d}
\int {\rm d}\bfr' \frac{ \bfr - \bfr' }{ |\bfr - \bfr' |^d} \cdot \bfF(\bfr') 
\end{align}
with $c_d=4\pi$ if $d=3$ and $c_d=2\pi$ if $d=2$. In practice, this
integration of the gradient can also be done efficiently in reciprocal
space, using Fast Fourier Transforms (FFT), from the force density discretized
on a grid (see Refs~\citenum{borgis_computation_2013, coles_computing_2019}).

\begin{figure}[ht!]
\includegraphics[width=0.98\columnwidth]{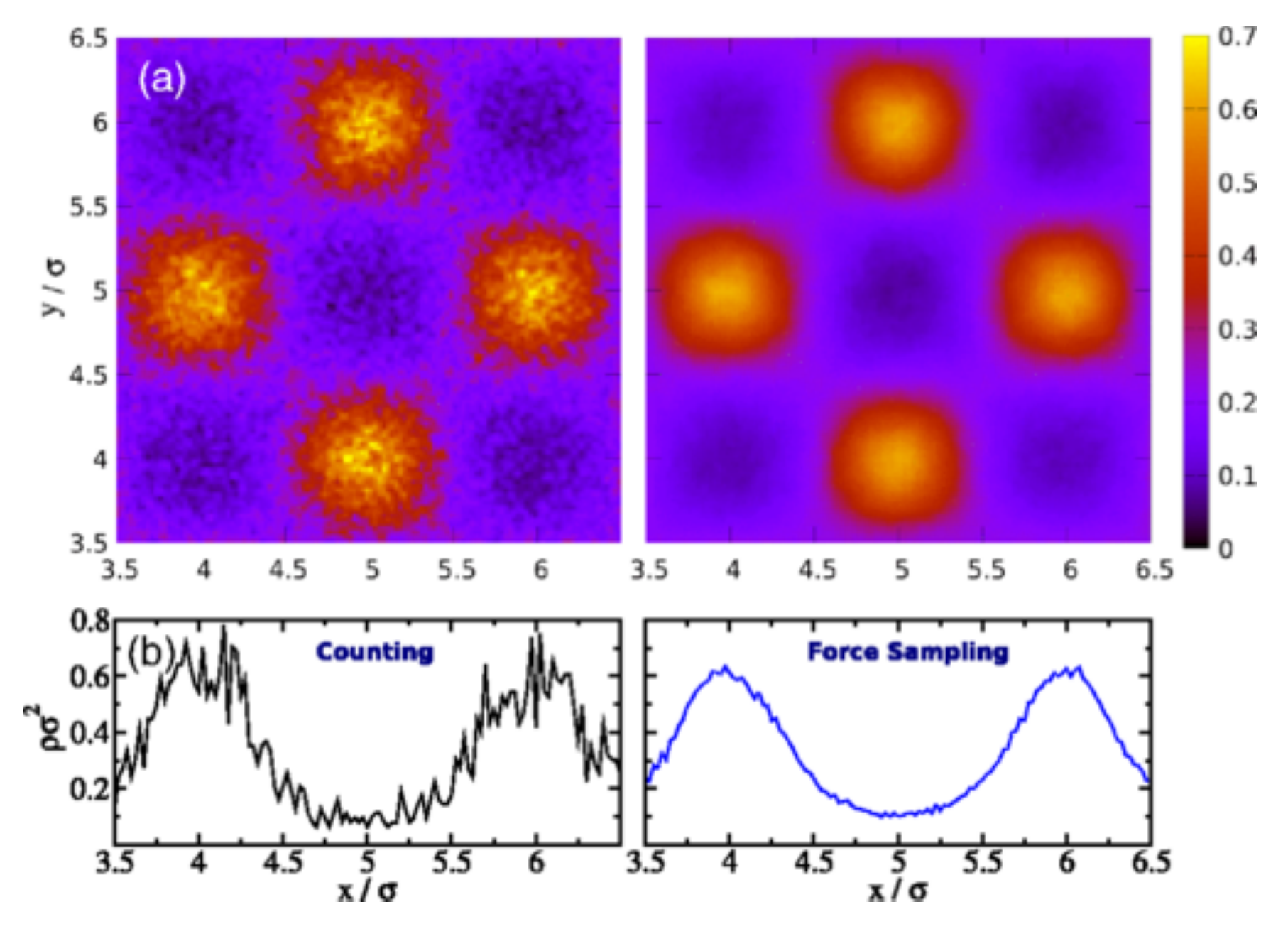}
\caption{
(a) 2D density for Lennard-Jones particles in an external potential 
$V_{ext}({\bf r}) = V_0 \sin(2\pi n_w x/L)\sin(2\pi n_w y/L)$
with $V_0=\epsilon$, the LJ energy, $n_w=5$ and $L=10\sigma$, with
$\sigma$ the LJ diameter. Simulations are performed with 25 particles and only
the central region of the box is shown. Results are obtained from $10^6$
configurations using histograms (left) and force sampling (right);
bins are squares of side length $0.025\sigma$. 
(b) Density profile as a function of $x$ at constant $y=5\sigma$
from histograms (left) and force sampling (right).
Reprinted with permission from de Las Heras and Schmidt,
\emph{Phys. Rev. Lett.} 2018, {\bf 120}, 218001. Copyright (2020) by the
American Physical Society.
}
\label{fig:density:2D}
\end{figure}

Figure~\ref{fig:density:2D} illustrates the benefit of the force sampling
approach on a 2D system of Lennard-Jones particles in an external
potential, from de Las Heras and Schmidt~\cite{de_las_heras_better_2018}. Both panels,
showing respectively a 2D density map and a 1D cut through the latter,
clearly demonstrate that using data from the same configurations from MC 
simulations and the same grid, the force-based estimator displays much 
less noise than the direct histogram approach. As expected from the discussion
of Section~\ref{sec:force}, the error with respect to the converged result with
many configurations diverges as the bin size decreases and behaves much better
with force scaling (see Ref.~\citenum{de_las_heras_better_2018} for a discussion
of the scaling on a 1D example, and Ref.~\citenum{coles_computing_2019} on a 3D
example). While force sampling provides a clear reduction of variance compared to
histograms for small bins, the benefit is less obvious for larger ones (a more
precise statement would have to be system specific), in particular because the
numerical integration of the gradient on a coarse grid also introduces some
discretization errors.

%%%%%%%%%%%%%%%%%%%%%%%%%%%%%%%%%%%%%
\subsection{Pitfalls and mitigation}
\label{sec:density:pitfalls}

\begin{figure}[ht!]
\includegraphics[width=0.98\columnwidth]{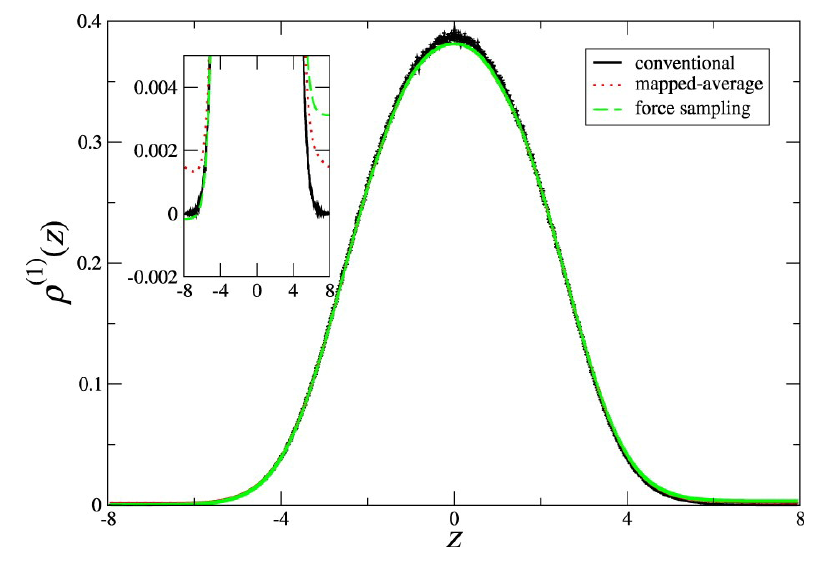}
\caption{
Density profile for a Lennard-Jones fluid in an external potential $\epsilon(z/\sigma)^2$, with $\sigma$
and $\epsilon$ the LJ diameter and energy. The results obtained from Monte Carlo
simulations are shown in units $\sigma=1$ with three estimators:
conventional histograms, force sampling and mapped-averaging (see
Eq.~\ref{eq:density1D:mappedaveraging} for the latter).
The inset shows the same data on a scale allowing a better visualization of the
tails of the distributions.
Reproduced from Purohit \emph{et al.}; \emph{Molecular Physics} 2019, {\bf 117}, 2822, with
permission of Taylor \& Francis Ltd.
}
\label{fig:density:1D}
\end{figure}

Another point deserving the attention of the reader is that the force sampling
approach may result in non-zero values of the density in regions where no
particles are present. This is illustrated on a 1D example in
Figure~\ref{fig:density:1D}, from Purohit \emph{et al.}\cite{purohit_force-sampling_2019},
which compares the conventional binning approach to force sampling
as described above and to mapped averaging, which is a more general framework
that can be applied to recover force sampling methods for density distributions 
(see below). All three methods provide similar
results everywhere, but the inset shows that they differ in the tails of the
distribution. While with histograms the density falls to zero when the external
potential diverges, this is not exactly the case for methods based on the force
(which, on the positive side, display a lower variance):
by integrating the density gradient as described above, the density starts from
0 on the left but integrates to a finite plateau on the right; mapped averaging
provides a symmetrized but non-vanishing result. Fortunately, this spurious
residual density, which comes from inaccuracies in the force density, can be 
reduced simply by increasing the number of configurations used to sample the latter.

Despite the above limitation (which also applies in higher dimensions), the
1D case also offers an example where the gradient can be integrated analytically.
The above-mentioned mapped averaging framework provides a way to introduce
approximate theoretical results (such as an \emph{a priori} estimate of a density)
and obtain an exact expression of the error in the theory~\cite{schultz_alternatives_2019, 
trokhymchuk_alternative_2019}. In the present 1D geometry, introducing 
a uniform density as a prior guess, Purohit \emph{et al.} obtained the following
expression~\cite{purohit_force-sampling_2019} 
\begin{align}
\label{eq:density1D:mappedaveraging}
\rho(z) &= \frac{N}{V} - \left\langle  \frac{1}{S} \sum_{i=1}^{N} 
\left( \frac{1}{2} - H( z - z_i ) - \frac{z_i - z}{L} \right) \beta f_{z,i}
\; ,
\right\rangle
\end{align} 
with $L$ the length of the system in the $z$ direction and $S=V/L$ 
the area of system in the lateral directions, $H$ the Heaviside function,
and $f_{z,i}$ the $z$-component of the force acting on particle $i$.
It is probably also possible, using a different prior estimate,
to recover from this formalism another expression obtained in 
Ref.~\citenum{mangaud_sampling_2020} from a different perspective,
namely to combine a position-based and a force-based estimator. To that end, one can 
introduce appropriate weights $w_N(z)$ and $w_f(z)$, where the subscripts refer to 
number and force, respectively, such that $w_f'(z)=\delta(z)-w_N(z)$
and that $w_f(z)$ vanishes when $|z|$ is large, in order to
write the 1D number density as:
\begin{align}
\label{eq:density1D:kernel}
\rho(z) &= \frac{1}{S}\left\langle  \sum_{i=1}^{N} \delta( z_i - z) \right\rangle
\nonumber \\
&= \frac{1}{S}\left\langle  \sum_{i=1}^{N} w_N( z_i - z) \right\rangle
-\frac{\beta}{S} \left\langle  \sum_{i=1}^{N} f_{z,i} w_f( z_i - z) \right\rangle
\end{align} 
The function $w_N(z)$ can be seen as a coarse-graining kernel for the
contribution of each particle to the number density and should
therefore vanish beyond a coarse-graining length $\xi$.
Families of estimators can be obtained by choosing various forms
for $w_N(z)$ and corresponding $w_f(z)$, and varying $\xi$
(see Ref.~\citenum{mangaud_sampling_2020} for an example of
weight functions and a discussion of the choice of $\xi$).
This provides a handle to mitigate the artefact of non-zero density in empty
regions predicted by integrating the force density.
The possible connection with mapped averaging can be hypothesized from
the observation that Eq.~\ref{eq:density1D:kernel} reduces to
Eq.~\ref{eq:density1D:mappedaveraging} (which is a particular case
of a more general expression~\cite{purohit_force-sampling_2019}) for 
$w_N(z)=1/L$ (which strickly speaking does not satisfy the criteria to be interpreted 
as a coarse-graining kernel, since every particle would contribute to the density everywhere).
In practice, the estimator defined in Eq.~\ref{eq:density1D:kernel}
can be computed efficiently by convoluting \emph{a posteriori} the
histogram-based estimators of the number and force densities with their
corresponding weight functions.

%%%%%%%%%%%%%%%%%%%%%%%%%%%%%%%%%%%%%
\subsection{Extensions: rigid bodies, generic densities}
\label{sec:density:extensions}

The force sampling approach to number density has recently been extended in
several directions by Coles \emph{et al.}~\cite{coles_computing_2019}. 
From the practical point of view, it is important to consider the case of
molecules described as rigid bodies using distance constraints,
which includes popular water models such as SPC/E or the TIP$n$P family. 
Even though the derivation is not straightforward, the final result is
particularly simple~\cite{coles_computing_2019}:
\begin{align}
\label{eq:densitygradientconstraints}
\nabla\rho(\bfr) &= \beta \bfF(\bfr)
= \beta
\left\langle  \sum_{i=1}^{N} \delta( \bfr_i - \bfr) \bff_i^*
\right\rangle_{constr.}
\; ,
\end{align} 
where $\bff_i^*$ is \emph{the sum of forces acting on particle $i$ and all
particles participating in a constraint with $i$} and where the 
the average is made over configuration satisfying all constraints.
In practice, the gradient of the water oxygen (resp. hydrogen) density is computed
by assigning the total force acting on each molecule to the position
of the O atom (resp. H atoms).

Another generalization was to consider other local quantities such as
the charge or polarization densities. Such an extension is trivial for
combinations of the type
\begin{align}
\label{eq:genericdensitydef}
A(\bfr) &= \left\langle 
\sum_{i=1}^{N} \delta( \bfr_i - \bfr) a_i \right\rangle_{constr.}
\end{align} 
when the microscopic property $a_i$ does \emph{not} depend on the coordinates
$\bfrN$, which is the case for the charge density. Following the same
derivation as for the number density, one readily obtains:
\begin{align}
\label{eq:genericdensitygradient}
\nabla A(\bfr) &= \beta
\left\langle  \sum_{i=1}^{N} \delta( \bfr_i - \bfr) a_i \bff_i^*
\right\rangle_{constr.}
\end{align} 
Unlike the charge density, the polarization density depends on the orientation
of the molecules, hence their atomic coordinates. For rigid polar molecules,
it is nevertheless possible to use a different set of coordinates, namely
the positions of their centers of mass ${\bf R}^{N_r}$ and orientations
$\bm{\Omega}^{N_r}$, with $N_r$ the number of rigid molecules. The result 
for the $\alpha$ component of the polarization density ${\bf P}(\bfr)$ is:
\begin{align}
\label{eq:polarizationdensitygradient}
\nabla P_\alpha(\bfr) &= \beta
\left\langle  \sum_{l=1}^{N_r} \delta( \bfR_l - \bfr) \mu_{l,\alpha} \bff_l^*
\right\rangle
\end{align} 
where the sum runs over molecules $l$ and $\mu_{l,\alpha}$
is the $\alpha$ component of their molecular dipole.
In that case there is no need to include constraints explicitly in the average
since there are no additional degrees of freedom.

\begin{figure}[ht!]
\includegraphics[width=\columnwidth]{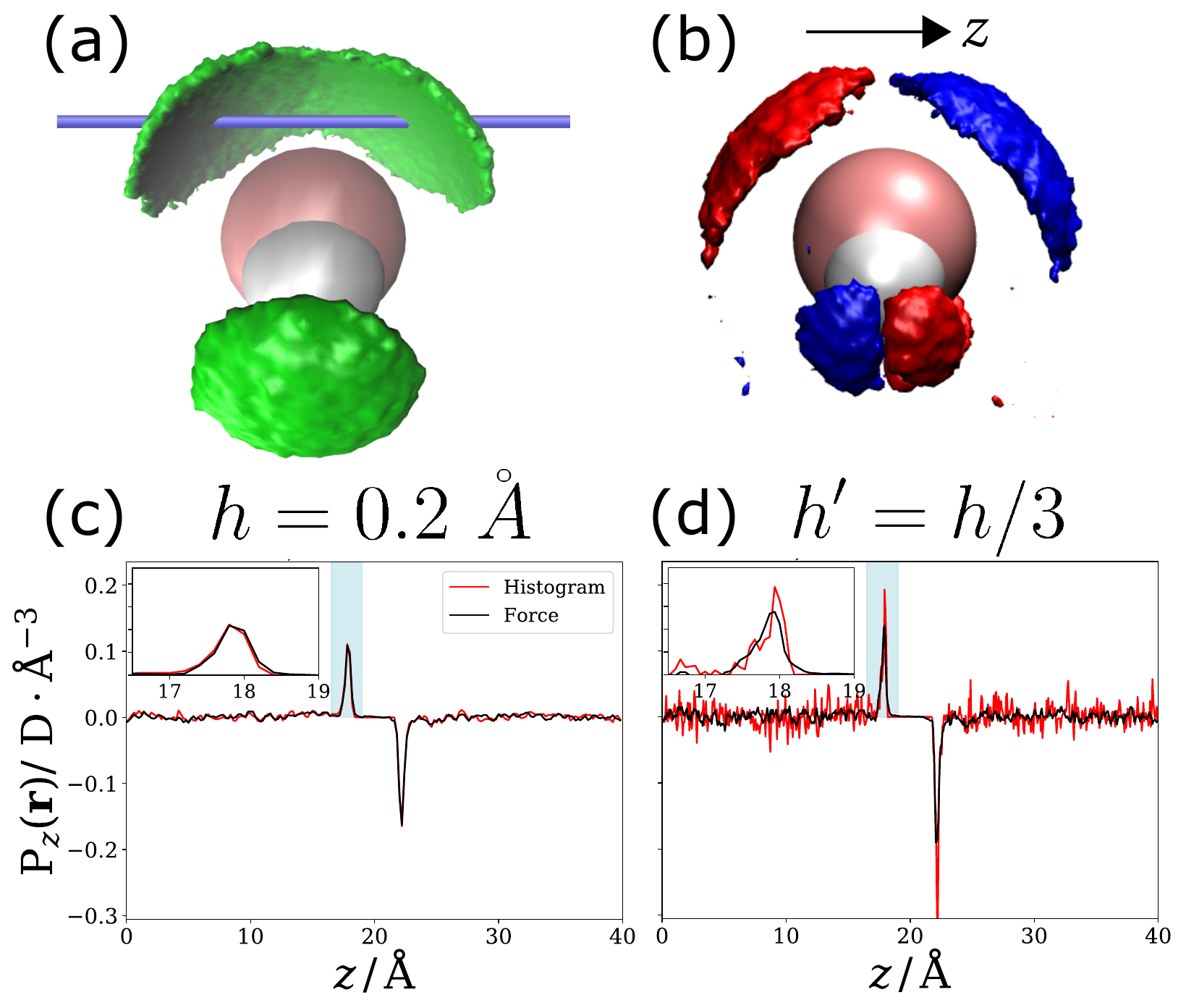}
\caption{
Water structure around a fixed water molecule, 
from simulations with the rigid SPC/E water model.
(a) Number density; the isosurface bounds the regions where the density
is greater than $0.07$~\AA$^{-3}$ and illustrates the position of water molecules
in the first solvation shell. The blue line indicates the position of
the axis along which the data of panels (c) and (d) are shown.
(b) $z$-component of the polarization density;
blue and red isosurfaces bound regions where the
$z$-component of the polarization density is less than $-0.035$~D~\AA$^{-3}$
and greater than +0.035~D~\AA$^{-3}$, respectively.
The remaining panels show the $z$-component of the polarization density for 
the single line of voxels illustrated in panel (a), using histograms (red)
or force sampling (black), using a grid spacing $h=0.2$~\AA\ (c)
or $h'=h/3$ (d). The insets are close-ups on the rising edge
highlighted in blue on the main figures.
Adapted from \emph{J. Chem. Phys.} 2019, {\bf 151}, 064124, with
the permission of AIP Publishing.
}
\label{fig:water}
\end{figure}

Figure~\ref{fig:water} illustrates results on the 3D organization of water,
around a fixed water molecule from simulations with the rigid SPC/E water model,
with results from Ref.~\citenum{coles_computing_2019}.
Panel~\ref{fig:water}a shows the regions of large number density of O atoms,
computed using Eq.~\ref{eq:densitygradientconstraints} on a grid and integrating the
gradient numerically using FFT. One can identify the position of the first solvation shell, 
with different basins corresponding to molecules donating (top) or receiving (bottom, 
only one is visible from this angle) H-bonds to/from the central molecule.
Panel~\ref{fig:water}b then shows the $z$ component of the polarization density,
computed using Eq.~\ref{eq:polarizationdensitygradient} and FFT to integrate the
gradient. This allows clarifiying the orientation of the molecular dipole of
water molecules in the different basins of panel~\ref{fig:water}a, consistent
with the donation/reception of H-bonds, and shows in particular that $P_z$ vanishes 
in the plane of the central molecule, as expected from the symmetry of the system.

In order to demonstrate the benefit of force sampling compared to histograms
to determine the 3D number and polarization densities, panels~\ref{fig:water}c
and~\ref{fig:water}d finally show a 1D trace through the 3D polarization density
along the blue line shown in panel~\ref{fig:water}a. Both methods display two
peaks of opposite sign in the vicinity of the molecule, corresponding to the
two basins for H-bond donating molecules of panel~\ref{fig:water}b. 
For a relatively large bin size, both methods give comparable results.
As the bin size decreases, the quality of the histogram-based estimator
deteriorates much faster than the force-based estimator: This is visible
in the larger amplitude of the noise far from the solute, as well as near the rising
edge close to the latter (see the insets) and the growing asymmetry between the
two sides of the molecule.

\begin{figure}[ht!]
\includegraphics[width=0.95\columnwidth]{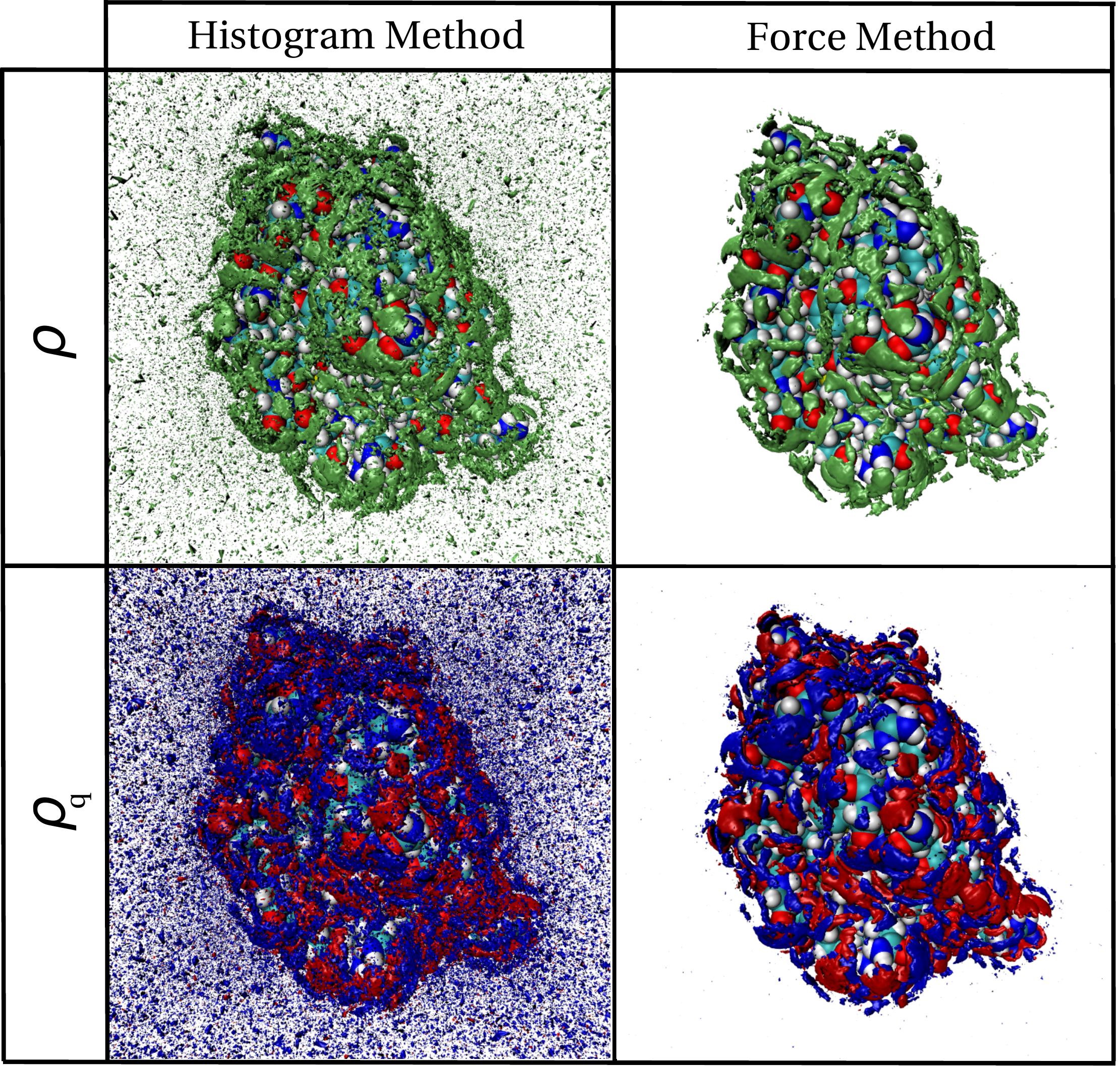}
\caption{
Isosurfaces for the number density $\rho$ (top) and the charge density 
$\rho_q$ (bottom) around a lysozyme protein in water, as obtained from 
histograms (left) or force sampling (right). 
The green surfarces bound the region where the number density
is greater than $0.1$~\AA$^{-3}$, while the blue (negative) and 
red (positive) surfaces bound regions where the magnitude of the charge density
exceeds $\pm0.1~e$~\AA$^{-3}$.
Reprinted from \emph{J. Chem. Phys.} 2019, {\bf 151}, 064124, with
the permission of AIP Publishing.
}
\label{fig:lysozyme}
\end{figure}

As a final illustration on a more complex system, Figure~\ref{fig:lysozyme}, 
also from Ref.~\citenum{coles_computing_2019}, compares the number and charge
densities around a small protein, lysozyme. Following Eq.~\ref{eq:genericdensitygradient}, 
the position and charge $q_i$ of each atom is considered when computing the
gradient, but the relevant force is the total force acting on the rigid
molecule. The advantage of force sampling over histograms is again clearly visible
in the reduction of the noise away from the solute, as well as in a better
resolved solvation structure near its surface. Among other features, the
positive and negative lobes of the charge density allow one to identify
molecules donating or receiving H-bonds.

%%%%%%%%%%%%%%%%%%%%%%%%%%%%%%%%%%%%%%%%%%%%%%%%%%%%%%%%%%%%%%%%%%%%%%%%%%
%%%%%%%%%%%%%%%%%%%%%%%%%%%%%%%%%%%%%%%%%%%%%%%%%%%%%%%%%%%%%%%%%%%%%%%%%%
\section{Radial distribution function}
\label{sec:rdf}

While the previous section only considered one-body densities, other statistical tools 
exist to quantify the structure of a system. We now turn to the force sampling
approach to radial distribution functions, which reflect correlations between particles.

%%%%%%%%%%%%%%%%%%%%%%%%%%%%%%%%%%%%%
\subsection{Virial-like estimator}
\label{sec:rdf:1}

Inspired by ideas for the electron density in Quantum Monte 
Carlo~\cite{assaraf_zero-variance_1999,assaraf_improved_2007,toulouse_zero-variance_2007}, 
Borgis \emph{et al.}~\cite{borgis_computation_2013} proposed an expression alternative 
to the definition Eq.~\ref{eq:rdfdef}. Starting from the latter, they used
Poisson's equation to rewrite $\delta(r)=-\frac{1}{4\pi}\Delta\frac{1}{r}$,
with $\Delta$ the Laplace operator. Integration by parts in the canonical
average, symmetrization beteween particles $i$ and $j$, and finally
Gauss's law then lead to:
\begin{align}
\label{eq:rdfforce}
g_{ab}(r) &=1+\frac{\beta}{4\pi}\frac{V}{N_a N_b}
\left\langle \sum_{i=1}^{N_a} 
\sideset{}{'}\sum_{j=1}^{N_b} 
\frac{1}{2}\left( \bff_i -  \bff_j \right)
\cdot \frac{\bfr_j-\bfr_i}{r_{ij}^3} H( r_{ij} - r) \right\rangle
\end{align}
with $H$ the Heaviside function. The origin of the forces $\bff_i$ 
and $\bff_j$ acting on the particles is again the gradient of the Boltzmann weight 
with respect to the particles positions, which comes from integrating by parts.
This expression displays naturally the separation between the ideal gas
contribution, which as described in Section~\ref{sec:force:binning} is detrimental
to the convergence of the histogram-based estimator, 
and a virial-like correction arising from interactions between particles and 
external potentials. It is worth emphasizing here that the above expression
holds even with many-body potentials, since no assumption of pairwise additivity
was needed in the derivation.

One can further note that it involves a Heaviside instead of a Dirac delta 
in Eq.~\ref{eq:rdfdef}. This results from the integration by parts in the canonical 
average, which can be seen as a way to ``integrate the gradient'' of $g$ analytically,
and has important practical consequences. While in the histogram approach each
pair only contributes to the estimate of the rdf in the bin corresponding to
$r=r_{ij}$ and symmetrically the estimate at $r$ only benefits from pairs 
such that $r_{ij}=r$, with Eq.~\ref{eq:rdfforce} each pair contributes to the
estimate of the rdf for all values $r\leq r_{ij}$ and symmetrically the estimate
of $g(r)$ benefits from all pairs separated by a distance larger than $r$.
In addition, bins are not necessary anymore and one can comptute the rdf with 
arbitrary resolution in $r$. 

\begin{figure}[ht!]
\includegraphics[width=0.9\columnwidth]{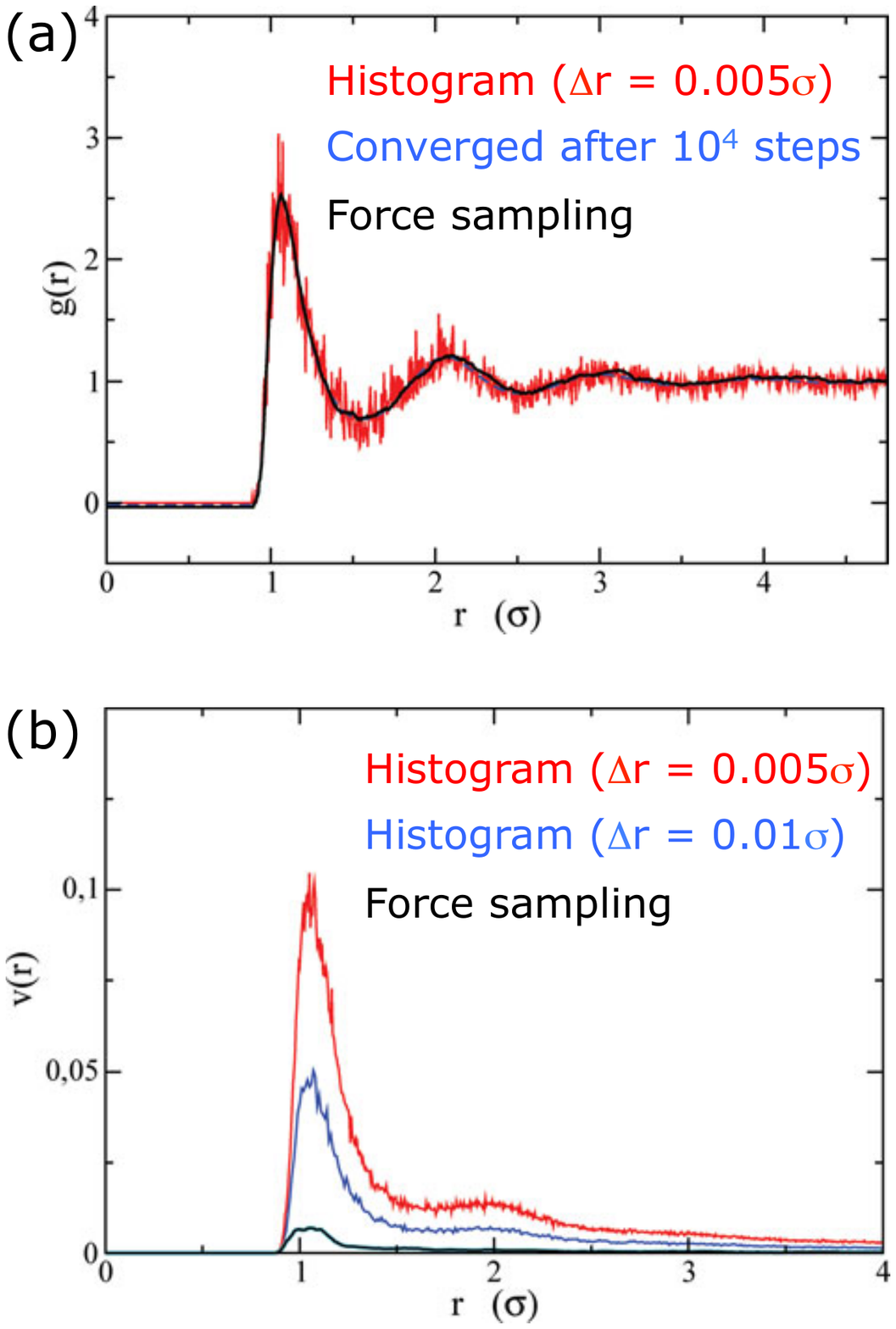}
\includegraphics[width=0.9\columnwidth]{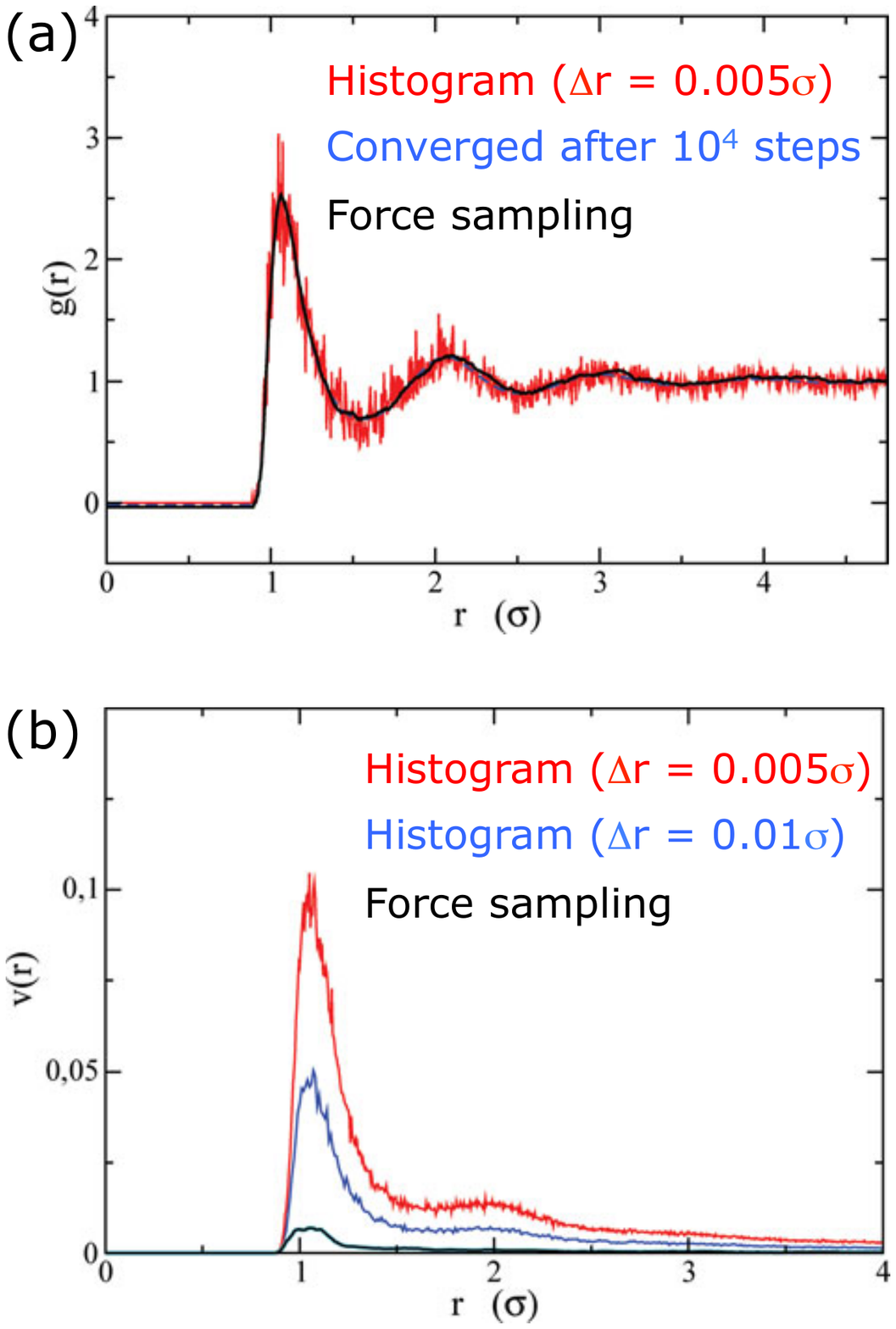}
\caption{
Radial distribution function (rdf) for a Lennard-Jones (LJ) fluid
at reduced density $\rho^*=0.8$ and reduced density $T^*=1.35$,
from a simulation with 864 particles.
(a) rdf obtained from a single microscopic configuration with 
histograms  (red) with a small bin width $\Delta r=0.005\sigma$,
with $\sigma$ the LJ diameter, and with force sampling (black);
the latter is indistinguishable from the converged result with histograms
averaged over $10^4$ configurations (blue). 
(b) Variance of the estimators, for each distance $r$, over $10^3$
configurations: histograms with $\Delta r=0.005\sigma$ (red) and $\Delta
r=0.01\sigma$ (blue), and force sampling (black).
Reproduced from Borgis \emph{et al.}, \emph{Molecular Physics} 2013, {\bf 111}, 3486, with
permission of Taylor \& Francis Ltd.
}
\label{fig:rdf}
\end{figure}

The benefit of the force sampling approach to estimate rdfs is illustrated in
Figure~\ref{fig:rdf}, which reproduces results from
Ref.~\citenum{borgis_computation_2013} for a Lennard-Jones fluid.
Panel~\ref{fig:rdf}a shows that the estimate obtained on a single configuration
with histograms (using a very small bin width) is much noisier than with force
sampling, the latter being indistinguishable from the converged histogram-based
estimate from $10^4$ configurations. Panel~\ref{fig:rdf}b then shows the
variance of the estimator over $N_{conf}=10^3$ configurations as a function of $r$,
defined by
\begin{align}
v(r) &= \frac{1}{N_{conf}}\sum_{k=1}^{N_{conf}} g_k(r)^2
- \left(\frac{1}{N_{conf}}\sum_{k=1}^{N_{conf}} g_k(r) \right)^2
\end{align}
The shape of $v(r)$ is the same for all methods, in particular it vanishes 
inside the core (where $g$ should vanish) and at long distance (where $g\approx1$).
However, the variance of the force-based estimator is significantly 
reduced compared to histograms, even more so that the bin size is small for the latter. 

%%%%%%%%%%%%%%%%%%%%%%%%%%%%%%%%%%%%%
\subsection{Alternative derivations and expressions}
\label{sec:rdf:2}

As mentioned earlier, Eq.~\ref{eq:rdfforce} can be seen as an integration of 
$g'(r)=dg/dr$ from $r\to\infty$ where $g=1$, leading to two difficulties. 
Firstly, in a finite box the limiting value of $g$ is never reached exactly.
In fact for simulations in the canonical ensemble, the plateau value 
differs from 1 by a $\mathcal{O}(N^{-1})$
correction~\cite{lebowitz_long-range_1961,belloni_finite-size_2018}.
Secondly, this expression does not guarantee \emph{a priori} that $g$ will vanish 
inside the core, which is the case for $r=0$ only if
\begin{align}
\frac{\beta}{4\pi}\frac{V}{N_a N_b}
\left\langle \sum_{i=1}^{N_a} 
\sideset{}{'}\sum_{j=1}^{N_b} 
\frac{1}{2}\left( \bff_i -  \bff_j \right)
\cdot \frac{\bfr_j-\bfr_i}{r_{ij}^3} \right\rangle
&= -1
\; .
\end{align}
In practice, this is almost the case (hence it is hard to see the small but
finite value in Figure~\ref{fig:rdf}a), but only approximately.
For systems with sufficiently hard repulsion at short range, one can 
enforce that $g(0)=0$ and integrate from $r=0$ to obtain an alternative expression
\begin{align}
\label{eq:rdfforce2}
g_{ab}(r) &=\frac{\beta}{4\pi}\frac{V}{N_a N_b}
\left\langle \sum_{i=1}^{N_a} 
\sideset{}{'}\sum_{j=1}^{N_b} 
\frac{1}{2}\left( \bff_i -  \bff_j \right)
\cdot \frac{\bfr_j-\bfr_i}{r_{ij}^3} H( r- r_{ij} ) \right\rangle
\; .
\end{align}
A similar expression was obtained for the two-body density by Purohit \emph{et al.} using 
mapped averaging (already discussed for the one-body density in 
Section~\ref{sec:density})~\cite{purohit_force-sampling_2019}.
Compared to Eq.~\ref{eq:rdfforce}, each pair now contributes to the estimate of the rdf 
for all values $r\geq r_{ij}$ and symmetrically the estimate of $g(r)$ benefits from all 
pairs separated by a smaller larger than $r$.

It is worth noting that another similar expression had in fact been
introduced earlier by Adib and Jarzynski~\cite{adib_unbiased_2005} 
as an unbiased estimator first for the density around a fixed solute, 
then for the rdf. Starting from the expression of a local (surface) quantity in terms
of a volume average, using Gauss's theorem and introducing an appropriate vector
field ${\bf u}$, they obtained an expression of the rdf (Eq.~20 of
Ref.~\citenum{adib_unbiased_2005}) valid for pairwise additive potentials
-- while the above expressions do not rely on such an approximation and
can be used also with many-body potentials such as in \emph{ab initio}
simulations. The vector field ${\bf u}$ introduced in this work is
similar to the $\nabla\frac{1}{r}=-\frac{\bf r}{r^3}$ in the above
expressions, but also accounts for the possible presence of a hard sphere solute
at the origin and introduces the largest sphere that fits in the simulation
box ($R_{max}=L/2$ with $L$ the box length). 
It might therefore be useful to consider a similar quantity in the
derivation of Ref.~\citenum{borgis_computation_2013}.

As a final remark on rdfs, we also mention that the idea of integrating by parts
in the canonical average can be pushed even further. In the same
study~\cite{borgis_computation_2013}, Borgis \emph{et al.} obtained by
performing a second integration by parts another expression of the rdf:
\begin{align}
\label{eq:rdfforcegradient}
g_{ab}(r) &=1+\frac{\beta}{4\pi}\frac{V}{N_a N_b}
\left\langle \sum_{i=1}^{N_a} 
\sideset{}{'}\sum_{j=1}^{N_b} 
\frac{1}{2}\left( \Phi_i +  \Phi_j \right)
\min(\frac{1}{r}, \frac{1}{r_{ij}}) \right\rangle
\end{align}
with $\Phi_i = \beta \bff_i^2 - \Delta_{\bfr_i} U$.
Unlike Eq.~\ref{eq:rdfforce} in which only the force enters, this one requires the Laplacian 
of the energy (trace of the force gradient) with respect to $\bfr_i$, which is usually
not computed in molecular dynamics simulations. In addition, preliminary
numerical tests in this study proved disappointing. However, this illustrates the
fact that there are many possibilities to obtain alternative estimators (even
though not always practically useful, at least immediately).

%%%%%%%%%%%%%%%%%%%%%%%%%%%%%%%%%%%%%%%%%%%%%%%%%%%%%%%%%%%%%%%%%%%%%%%%%%
%%%%%%%%%%%%%%%%%%%%%%%%%%%%%%%%%%%%%%%%%%%%%%%%%%%%%%%%%%%%%%%%%%%%%%%%%%
\section{Local transport coefficients}
\label{sec:transport}

\begin{figure}[ht!]
\includegraphics[width=0.9\columnwidth]{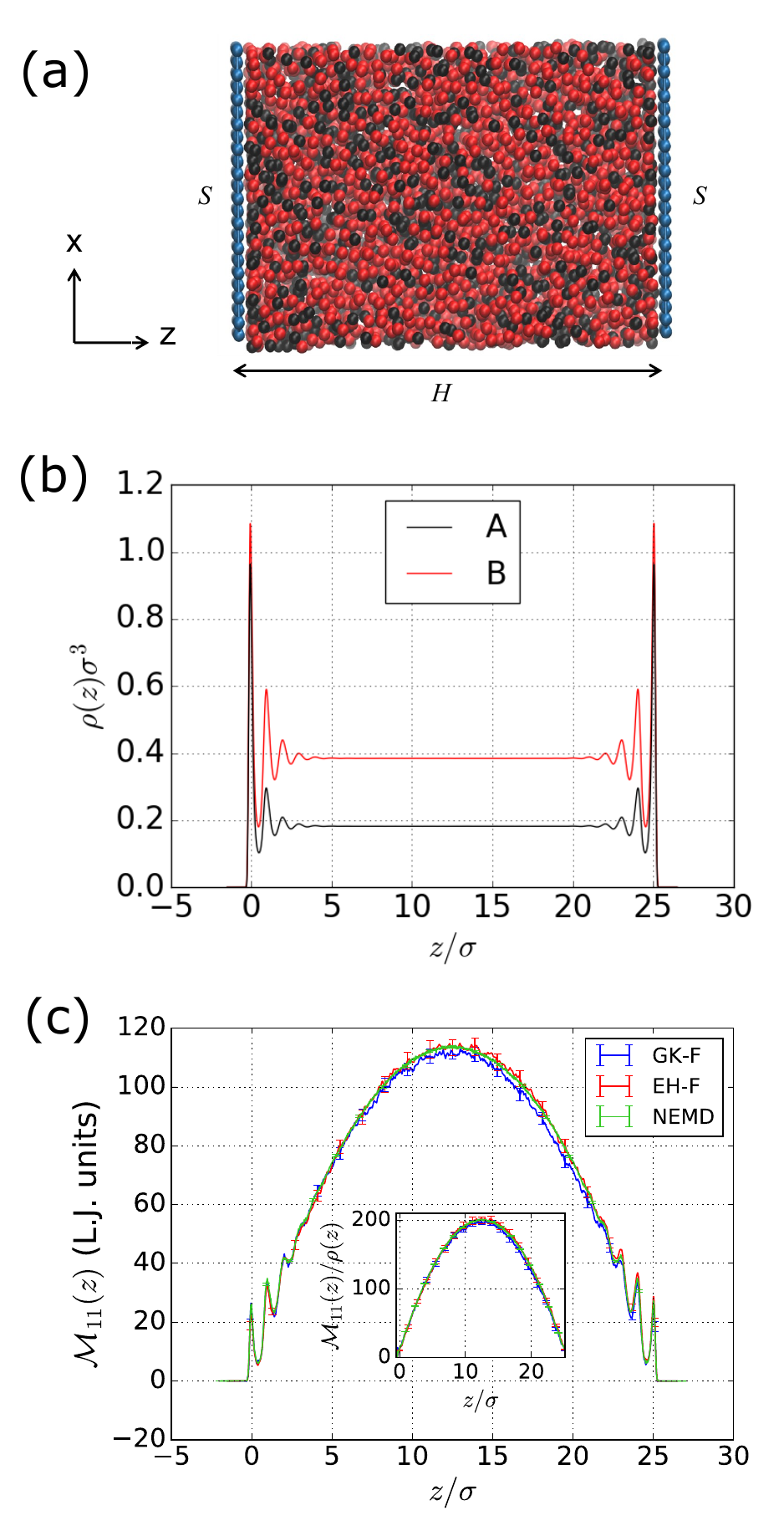}
\caption{
(a) Binary Lennard-Jones (LJ) fluid confined between walls consisting of LJ
particles. The solute A (black) and solvent B (red) are identical but the solute
has a stronger interaction with the wall.
(b) Density profiles; the position and density are
given in units $\sigma$ and $\sigma^{-3}$, respectively, with $\sigma$ the
LJ diameter common to all species.
(c) Mobility profile $\mathcal{M}_{11}(z)$ describing the linear response of the
total flux to a pressure gradient (see Eq.~\ref{eq:mobility}), obtained
from equilibrium simulations using force sampling with Green-Kubo (GK) and
Einstein-Helfand (EH), as well as from non-equilibrium molecular dynamics
(NEMD). The inset shows that $\mathcal{M}_{11}(z)/\rho(z)$ exhibits the
parabolic shape expected from continuum hydrodynamics for the velocity profile
in this case (Poiseuille flow).
Adapted from \emph{J. Chem. Phys.} 2020, {\bf 153}, 044125 , with
the permission of AIP Publishing.
}
\label{fig:transport}
\end{figure}

Beyond structural properties, the force sampling strategy
has also been explored recently for the computation of local transport coefficients
in confined fluids~\cite{mangaud_sampling_2020}, such as the system illustrated 
in Figure~\ref{fig:transport}a. Due to the external potential from the walls, 
the fluid adopts a layered structure, shown in Figure~\ref{fig:transport}b.
In the presence of an external perturbation parallel to the walls, such as a pressure 
gradient $-\nabla P$ or a chemical potential gradient $-\nabla\mu$ 
(or an electric field $-\nabla\psi$, \emph{e.g.} for an electrolyte
solution) the various components of the fluid will respond differently
and the local steady-state flux of each species will depend on the type of 
perturbation. For sufficiently small perturbations, the response is linear and
fully characterized by a \emph{mobility matrix} $\mathcal{M}(z)$ relating local
fluxes to the external forces. For the binary fluid of Figure~\ref{fig:transport}, 
this can be written as
\begin{equation}
\left(\begin{array}{c} q(z) \\ j_A(z) - c_A^* q(z)  \end{array} \right) 
= \mathcal{M} (z)\left(\begin{array}{c} -\nabla P \\ -\nabla \mu  \end{array} \right)
\label{eq:mobility}
\;,
\end{equation}
where $q$ and $j_A$ are the local volume fluxes and solute fluxes, respectively,
and $c_A^*$ is a reference concentration related to the composition in the bulk
part of the fluid. For example, the element $\mathcal{M}_{11}(z)$ of this
$2\times2$ matrix predicts the flux of all species under a pressure gradient
(Poiseuille flow), while $\mathcal{M}_{12}(z)$ corresponds to diffusio-osmosis.

%%%%%%%%%%%%%%%%%%%%%%%%%%%%%%%%%%%%%
\subsection{Mobility profiles from equilibrium molecular dynamics}

For illustration purposes, we will only consider here $\mathcal{M}_{11}(z)$, but
the other cases can be found in Ref.~\citenum{mangaud_sampling_2020}.
Using linear response theory, it is possible
to derive expressions of the mobility profiles $\mathcal{M}_{ij}(z)$
as integrals of correlation functions (Green-Kubo approach, GK)
\begin{align}
\mathcal{M}_{11}^{GK}(z)&= \frac{  V }{ k_B T } \int_{0}^{+\infty} {\rm d}t\,
C_{qQ}(t,z)
  \label{eq:GK}
\; ,
\end{align}
with $C_{qQ}(t,z)=\left\langle q(z,t)Q(0) \right\rangle$
is a time-correlation function (for each $z$) between the local volume flux
\begin{align}
q (z,t) &=  \frac{H}{N} \sum \limits_{i=1}^{N} v_{x,i}(t) \delta (z_i(t)-z) 
\; ,
\label{eq:q_z} 
\end{align}
with $v_{x,i}$ the $x$-component of the velocity of particle $i$,
$N=N_A+N_B$ the total number of particles 
and $H$ the distance between the wall, and the global flux averaged
over the fluid slab
\begin{equation}
  Q(t)  = \frac{1}{H} \int_0^H {\rm d}z\, q(z,t) = \frac{1}{N} \sum
\limits_{i=1}^{N} v_{x,i}(t) 
  \label{eq:Qoft}
\; .
\end{equation}
Alternatively, the integrals can be computed as (Einstein-Helfand approach, EH) 
\begin{equation}
\mathcal{M}_{11}^{EH}(z)=\frac{V}{k_B T} 
\lim_{t \rightarrow +\infty} %\frac{K_{11}(t,z)}{2t} 
\frac{\left\langle \int_0^{t} {\rm d}t''\, q(z,t'') \int_0^{t} {\rm
d}t' Q(t') \right\rangle}{2t}
\; .
\label{eq:EH}
\end{equation}

%%%%%%%%%%%%%%%%%%%%%%%%%%%%%%%%%%%%%
\subsection{Force sampling for time correlation functions}

Since the time correlations are defined as canonical averages of observables
involving a Dirac delta, one can follow the ideas developed for static properties. 
From the definition
\begin{align}
C_{qQ}(t,z) &= \left\langle 
Q(0) \frac{H}{N} \sum \limits_{i=1}^{N} v_{x,i}(t) \delta (z_i(t)-z) 
\right\rangle
\; ,
\label{eq:c112}
\end{align}
one can introduce a force-weighted observable,
\begin{equation}
F_{qQ}(t,z) = \left\langle 
Q(0) \frac{H}{N} \sum \limits_{i=1}^{N} v_{x,i}(t) f_{z,i}(t)\delta (z_i(t)-z) 
\right\rangle
\; ,
\label{eq:c_f}
\end{equation}
and form, in the spirit of Eq.~\ref{eq:density1D:kernel}, the mixed estimator
\begin{align}
\tilde{C}_{qQ}(t,z) 
= &\int_0^H {\rm d}z'\, \left[ w_N(z'-z) C_{qQ}(t,z')  -  w_f(z'-z) \beta
F_{qQ}(t,z') \right]
\nonumber \\
= & (w_N*C_{qQ})(t,z)  +  \beta (w_f *F_{qQ})(t,z)
\; ,
\label{eq:c_f_sampl}
\end{align}
where the convolution products are in space only, not time.
A similar construction can be followed for the EH approach.

Figure~\ref{fig:transport}c compares the predictions from equilibrium MD
simulations with the GK and EH approaches, both using the mixed estimators
introduced above, for the $M_{11}(z)$ element of the mobility matrix,
corresponding to a Poiseuille flow. The results are found to be in excellent 
agreement with the more direct non-equilibrium MD approach. The inset further
shows that when considering the velocity profile, \emph{i.e.} the flux divided
by the density, one recovers the parabolic shape expected from continuum
hydrodynamics. The fact that the NEMD results are recovered validates the
relevance of the approach. This is particularly important because the derivation
leading to the exact result for the density is complicated for time-correlation 
functions by the fact that observables are considered at two times:
The canonical average corresponds to points in phase space at the initial time 0,
where the global flux is also considered, while the local fluxes are computed
from the positions and velocities at subsequent time $t$ (it is possible to
write a symmetric expression). The integration by parts over the initial
positions $z_i(0)$ then introduces an additional term, $\left\langle Q(0)
\frac{H}{N} \sum_{i=1}^N \frac{\partial v_{x,i}(t) }{\partial
z_i(0)} w_f(z_i-z) \right\rangle$,
which involves the derivative of the $x$ component of the velocity at time
$t$ with respect to the initial position in the $z$ direction.
Although this term might not vanish in principle, it would be very difficult 
to evaluate from the trajectories. It was found numerically that it
was sufficient to neglect it, but a formal derivation would be desirable.

The improvement of the forced-based estimator compared to
histograms is discussed in detail in Ref.~\citenum{mangaud_sampling_2020},
together with results for the other elements of the mobility matrix.
We only emphasize here that such an improvement is essential, 
because determining the mobility profiles is 
computationally much more demanding than static properties: It requires
converging, \emph{for all positions}, the time correlation functions with
sufficient accuracy to compute the integral in Eq.~\ref{eq:GK} (GK)
or the slope in Eq.~\ref{eq:EH} (EH). To conclude this discussion of local
transport properties, we note that for the determination of a single mobility
profile NEMD remains more efficient than the above equilibrium MD route.
However, the latter can provide all elements of the mobility matrix
simultaneously, whereas NEMD requires separate simulations with the
different perturbations, and to consider different magnitudes in order to check
the validity of the linear response. Therefore, one can expect that the
equilibrium approach will be particularly helpful for multicomponent systems,
\emph{e.g.} for the response of electrolyte solutions to pressure, chemical
potential or electric potential gradients (hence a $3\times3$ mobility matrix). 
The force based estimator, with a reduced variance compared to histograms, 
will contribute to decreasing the number and length of trajectories necessary 
to obtain reliable results.

%%%%%%%%%%%%%%%%%%%%%%%%%%%%%%%%%%%%%%%%%%%%%%%%%%%%%%%%%%%%%%%%%%%%%%%%%%
%%%%%%%%%%%%%%%%%%%%%%%%%%%%%%%%%%%%%%%%%%%%%%%%%%%%%%%%%%%%%%%%%%%%%%%%%%
\section{Discussion}
\label{sec:discussion}

In the previous sections, we have mentioned several approaches 
arriving from different perspectives to similar expressions 
for estimators with a reduced variance compared to histograms, 
which involve the forces acting on the particles. 
The examples making use of the Poisson equation to rewrite the Dirac 
delta and/or using Gauss's theorem involve integration by parts
in the definitions as canonical averages, very much in the spirit
of the Yvon theorem for averages of the derivative of an arbitrary function of the 
particle coordinates $A(\bfrN)$ with respect to the position $z_i$ of one 
particle~\cite{hansen_theory_2013}:
\begin{align}
\label{eq:yvon}
\left\langle \frac{ \partial A(\bfrN) }{\partial z_i} \right\rangle
&= \beta \left\langle A(\bfrN) \frac{ \partial U(\bfrN) }{\partial z_i} \right\rangle
= -\beta \left\langle A(\bfrN) f_{z,i} \right\rangle
\; .
\end{align}

They are also directly related to the notion of potential of mean force (PMF),
as discussed for the rdf in Ref.~\citenum{borgis_computation_2013} or the derivation of
Ref.~\citenum{coles_computing_2019} for the density gradient in the presence of
constraints, which makes use of the fundamental result of Ciccotti \emph{et al.}
for the mean force associated with generic collective variables~\cite{ciccotti_blue_2005}, 
by choosing the 3 spatial coordinates as collective variables. 
The same strategy could be used to extend Eq.~\ref{eq:rdfforce} in 
the presence of constraints, using the distance between particles as 
collective variable. One should keep in mind, however, that 
the force density $\bfF(\bfr)=k_BT\nabla\rho(\bfr)$ 
(see Eq.~\ref{eq:densitygradient}) and the mean force, derived from the PMF,
differ by a factor $\rho(\bfr)$. A more detailed discussion of this difference
and the practical consequences for their computation in molecular simulations,
in the case of the rdf, can be found in Ref.~\citenum{borgis_computation_2013}.

The separation between the ideal gas reference and the contribution from
interactions and external potentials is natural in the framework of mapped
averaging, which provides the exact correction to a theory (\emph{e.g.} uniform
density). This idea of using a reference distribution, leading to estimators
involving forces, had also been proposed by Basner and Jarzynski to compute binless 
PMFs, with an illustration on an angular distribution~\cite{basner_binless_2008}, 
or Zhang and Ma to compute rdfs or angular distributions~\cite{zhang_estimating_2012}.
One could envision for example to apply it also in the case of the local
transport properties illustrated in Section~\ref{sec:transport}, using the
theoretical prediction from continuum hydrodynamics as reference.

Since in molecular dynamics simulations the forces are computed to generate the 
trajectory, the force-based estimator does not entail a larger 
computational cost compared to histograms. As a result, one can only recommend
the use force-based estimators to compute local densities or radial distribution
functions. For Monte Carlo, the additional cost of computing the forces in addition 
to the energy should be compared to that of generating enough configurations to achieve 
the same variance reduction with histograms, which depends in particular on the bin width.
Another issue related to the computation of forces is the case of particles described 
by hard cores. Indeed, the expressions reported here cannot be used directly.
However, the effect of excluded volume can be treated separately from other
interactions, which allows to apply the ideas of force sampling even in this
case~\cite{adib_unbiased_2005,trokhymchuk_alternative_2019}.

%%%%%%%%%%%%%%%%%%%%%%%%%%%%%%%%%%%%%%%%%%%%%%%%%%%%%%%%%%%%%%%%%%%%%%%%%%
%%%%%%%%%%%%%%%%%%%%%%%%%%%%%%%%%%%%%%%%%%%%%%%%%%%%%%%%%%%%%%%%%%%%%%%%%%
\section{Summary and future challenges}
\label{sec:conclusion}

Even though the determination of local properties, such as densities or radial
distribution functions, remains one of the most standard goals of molecular
simulation, it seems that the community still essentially relies on
histogram-based algorithms to estimate them. The main objective of the present
work was to highlight recent developments of alternative approaches leading, 
from different perspectives, to estimators with a reduced variance compared to
conventional binning. They all make use of the force acting on the particles, 
in addition to their position, and allow to focus on the non-trivial part
of the problem in order to alleviate (or even remove in some cases) the
catastrophic behaviour of histograms as the bin size decreases.
The corresponding computational cost is negligible for molecular dynamics simulations,
since the forces are already computed to generate the configurations, and the
benefit of reduced-variance estimators will be even larger when the cost of
generating the latter is high, in particular with \emph{ab initio} simulations.
The force sampling approach may result in spurious residual non-zero values
of the density in regions where no particles are present, but strategies are
available to mitigate this artefact.

Up to now, the variance reduction via force sampling has mainly been
demonstrated numerically -- even though the methods themselves were of course
derived as rigorously as possible and good reasons were proposed to
explain the observations. Since variance reduction is a field of
research by itself, computational physicists, chemists and biologists 
would certainly benefit from insights from the Mathematics community
in order to obtain analytical predictions of the gain to be expected with respect to
binning, in particular the scaling with the bin size (when the force density is
first sampled on a grid, then integrated numerically) and the number of configurations. 
Other directions could include improved or optimal methods to ``integrate the gradient'', 
in order to reconstruct the density from the force density, in particular in 3D,
or the formal derivations in the case of transport (see the discussion of the
``missing term'' in Section~\ref{sec:transport}), including in the presence of
constraints for rigid bodies~\cite{ciccotti_holonomic_2018}.

The potential of these force sampling strategies could also be investigated
for other thermodynamic conditions, \emph{e.g.} the isothermal-isobaric ($NPT$)
or the grand-canonical ($\mu VT$) ensembles, or for quantities derived from the
ones considered here, \emph{e.g.} to improve the convergence of Kirkwood-Buff
integrals, which involve rdfs~\cite{kruger_kirkwoodbuff_2013,ganguly_convergence_2013}.
Other local properties might also benefit from the ideas behind force sampling,
in particular the local stress tensor~\cite{wajnryb_uniqueness_1995,lion_computing_2012}.
However, one should keep in mind that this might require quantities that are
not computed in typical simulations, such as the force gradient, and that the
additional computational cost should not exceed that of reducing the
variance by the same amount using histograms simply by generating more
configurations. Another possible direction for the short-term development
of force sampling, in particular following the approach of
Ref.~\citenum{coles_computing_2019} based on the results of
Ref.~\citenum{ciccotti_blue_2005}, is the extension to non-cartesian coordinates
and/or multidimensional collective variables, which can be used to characterize
the 3D structure of liquids (see \emph{e.g.} Ref.~\citenum{zhang_revealing_2020} for
a recent example). 

Finally, it would be interesting explore the connections with other simulation contexts
where the strategies discussed in the present work could also provide alternative 
expressions for observables of interest. For example, improved estimators involving 
the force have already been proposed in path integral molecular dynamics 
to estimate thermodynamic quantities, based on a virial form (see \emph{e.g.} 
Refs.~\citenum{glaesemann_improved_2002,korol_dimension-free_2020}).
In the other direction, the ability to reduce errors in the computation of
densities/PMFs would also be useful for coarse-graning strategies 
based \emph{e.g.} on Boltzmann inversion~\cite{lyubartsev_calculation_1995} 
or on the relative entropy~\cite{chaimovich_coarse-graining_2011},
or for simulations with adaptive resolution introducing a thermodynamic force
density in the hybrid region~\cite{fritsch_adaptive_2012}.
As for dynamic properties, beyond the equilibrium route presented in
Section~\ref{sec:transport}, it could be useful to extend the force-sampling
approach to non-equilibrium steady state, as suggested in
Ref.~\citenum{de_las_heras_better_2018}, or even to path sampling
methods~\cite{garrahan_first-order_2009}.
We hope that the present perspective will motivate others to embark on
this promising path.

%%%%%%%%%%%%%%%%%%%%%%%%%%%%%%%%%%%%%%%%%%%%%%%%%%%%%%%%%%%%%%%%%%%%%%%%%%
%%%%%%%%%%%%%%%%%%%%%%%%%%%%%%%%%%%%%%%%%%%%%%%%%%%%%%%%%%%%%%%%%%%%%%%%%%

\section*{Data availability}

Data sharing is not applicable to this article as no new data were created or
analyzed in this study.

%%%%%%%%%%%%%%%%%%%%%%%%%%%%%%%%%%%%%%%%%%%%%%%%%%%%%%%%%%%%%%%%%%%%%%%%%%
%%%%%%%%%%%%%%%%%%%%%%%%%%%%%%%%%%%%%%%%%%%%%%%%%%%%%%%%%%%%%%%%%%%%%%%%%%

%

%%%%%%%%%%%%%%%%%%%%%%%%%%%%%%%%%%%%%%%%%%%%%%%%%%%%%%%%%%%%%%%%%%%%%%%%%%
\begin{acknowledgments}
The author wishes to express his gratitude to many colleagues  
for collaborations, discussions and suggestions of literature on related work, 
in particular: Daniel Borgis, Rodolphe Vuilleumier, Roland Assaraf,
Samuel Coles, Etienne Mangaud, Daan Frenkel, Antoine Carof, Sara Bonella, 
Giovanni Ciccotti, David Kofke, Vincent Krakoviack, Thomas F. Miller III, Glen Hocky, 
and Juan P. Garrahan. He acknowledges financial support from the
Ville de Paris (Emergences, project Blue Energy),
the H2020-FETOPEN project NANOPHLOW (grant agreement No. 766972) 
from the European Research Council (ERC) under the European Union's Horizon 2020
research and innovation programme (grant agreement No. 863473).
\end{acknowledgments}

\end{document}